\documentclass[reprint, amsmath,amssymb,aps, prb,superscriptaddress]{revtex4-1}
\usepackage{amsfonts,amssymb}
\usepackage{graphicx}
\usepackage{hyperref}    
\usepackage{bm}          
\usepackage{natbib}
\usepackage{soul} 
\begin{document}

\title{Theory of a field effect transistor based on semiconductor nanocrystal array}

\author{K.V. Reich}
\email{Reich@mail.ioffe.ru}
\affiliation{Fine Theoretical Physics Institute, University of Minnesota, Minneapolis, MN 55455, USA}
\affiliation{Ioffe Physical-Technical Institute, St. Petersburg, 194021, Russia}
\author{Tianran Chen}
\affiliation{Fine Theoretical Physics Institute, University of Minnesota, Minneapolis, MN 55455, USA}
\author{B.I. Shklovskii}
\affiliation{Fine Theoretical Physics Institute, University of Minnesota, Minneapolis, MN 55455, USA}

\begin{abstract}
We study  the surface conductivity of a field-effect transistor (FET) made of  periodic array of spherical semiconductor nanocrystals (NCs). We show that electrons introduced to  NCs by the gate voltage occupy one or two layers of the array. Computer simulations and analytical theory are used to study the array screening and corresponding evolution of electron concentrations of the first and second layers with growing gate voltage. When  first layer  NCs have two electrons per NC the quantization energy gap between its $1S$ and $1P$ levels induces occupation of $1S$ levels of   second layer NCs. Only at a larger gate voltage electrons start leaving $1S$ levels of second layer NCs and filling $1P$ levels of  first layer NCs.  By substantially larger gate voltage, all the electrons vacate the  second layer and move to   $1P$ levels of   first layer NCs. As a result of this nontrivial evolution of the two layers  concentrations, the surface conductivity of FET non-monotonically depends on the gate voltage. The same evolution of electron concentrations leads to non-monotonous behaviour of the differential capacitance.  
\end{abstract}



\date{\today}

\maketitle

\section{Introduction} \label{sec:intro}


In the recent years, there has been growing interest in investigation of the semiconductor nanocrystals (NCs) due to their size-tunable optical and electronic properties~\cite{Fernee_Tamarat_Lounis_2014, Lee_Lee_Song_Hiramoto_2013,Lee_Kovalenko_Shevchenko_2010,Dondapati_Ha_Pradhan_2013,Kim_Kotov_2013}. Like in bulk semiconductors, adding charged carriers is critical for NC solids, which would otherwise be electrically insulating. 
There are a few theoretical works on the conductivity of impurity doped arrays of NCs~\cite{Conductivity_Brian_Shklovskii, Shabaev_Efros_Efros_2013,Cachopo_Poncet_le_Carval_2012, pen_Nelson_Vanmaekelbergh_2007}. However, due to intrinsic difficulties of doping of NCs by impurity atoms~\cite{NorrisScience08}, so far there are only a limited number of experimental works with successful bulk doping~\cite{Kang_Sahu_Frisbie_Norris_2013,g_Winterer_Frisbie_Norris_2012,attner_Millo_Rabani_Banin_2011}. On the other hand,  introducing carriers via field-effect is a more successful approach~\cite{lenko_Huang_Chung_Talapin_2011,Kang_Sahu_Frisbie_Norris_2013,iu_Crawford_Hemminger_Law_2013, un_Malliaras_Hanrath_Wise_2012,ki_Supran_Bawendi_Bulovic_2012,Robel_Lee_Pietryga_Klimov_2013}. The whole recent discussion of record band-like mobilities of NC arrays is centered around recent field-effect data \cite{Talapin_Magnet,Talapin_2013_high_mobility,iu_Crawford_Hemminger_Law_2013,Guyot-Sionnest_2012,Murray_Bandlike}.

\begin{figure}[t]
\includegraphics[width=\linewidth]{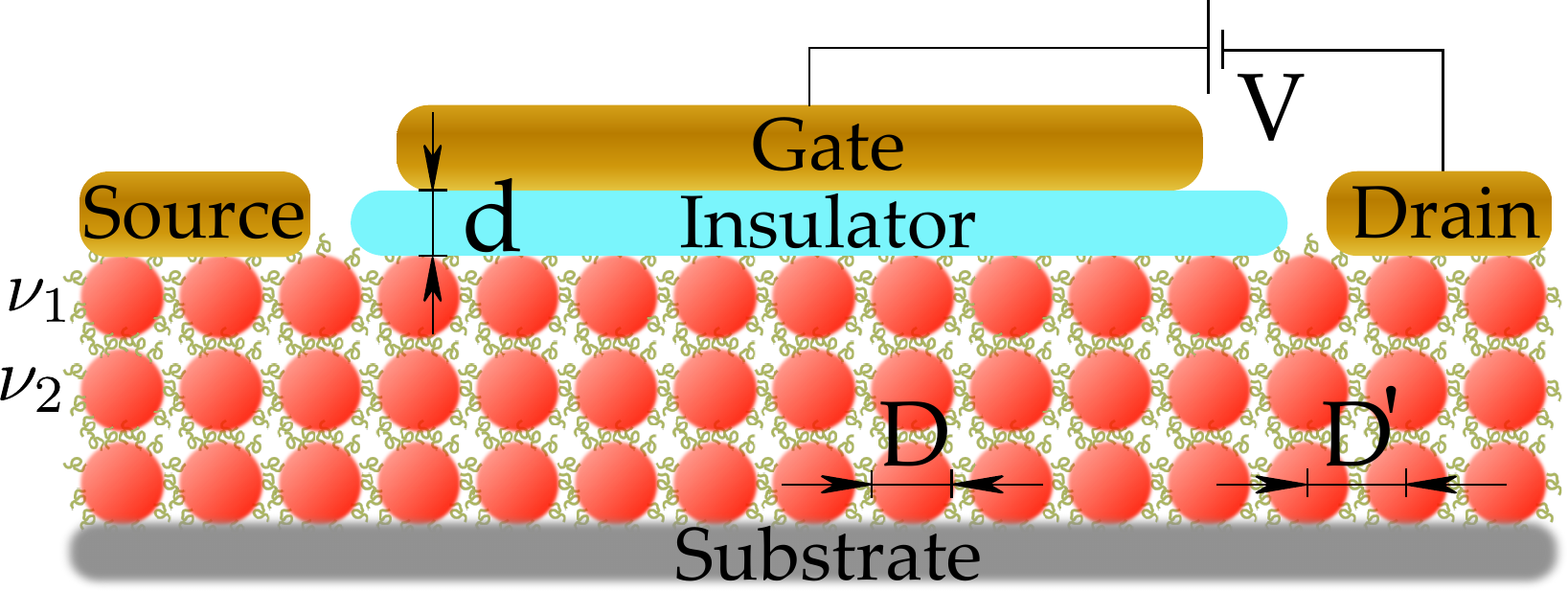}
\caption{(Color  online) Field-effect transistor based on array of spherical semiconductor NCs  with diameter $D$ and lattice constant $D'$. Each NC is coated by a thin layer of insulating ligands (curvy lines) that maintains a separation  $D' - D$ between NCs.   The distance between NCs in the first layer and the gate is equal to $d$. Applied gate voltage $V$ induces $\nu_1,\nu_2$ electrons per NC  in the first  and second  layer respectively, which  screen the exterior electric field of  the gate.}
\label{fig:FET}
\end{figure}

Meanwhile, there is no theory of  conductivity of field-effect transistor (FET) based on NC array beyond those which are based on mean field theories~\cite{etsch_Zhao_Kershaw_Rogach_2013,iu_Crawford_Hemminger_Law_2013}. Debye-H\"uckel-like estimates (see below) show that electrons introduced to a NC array  by the gate voltage occupy one or two  layers of the array closest to the gate  (see  Fig.~\ref{fig:FET}). Below we suggest a more detailed, non-mean-field theory of the array screening which allows for discreteness of layers in the array, discreteness of  NC charge and large quantization gap between $1S$ and $1P$ levels of a NC.  It results in non-trivial evolution of electron concentrations and low temperature conductivities of the first and second layers with growing gate voltage.

\begin{figure}[t]
\includegraphics[width=\linewidth]{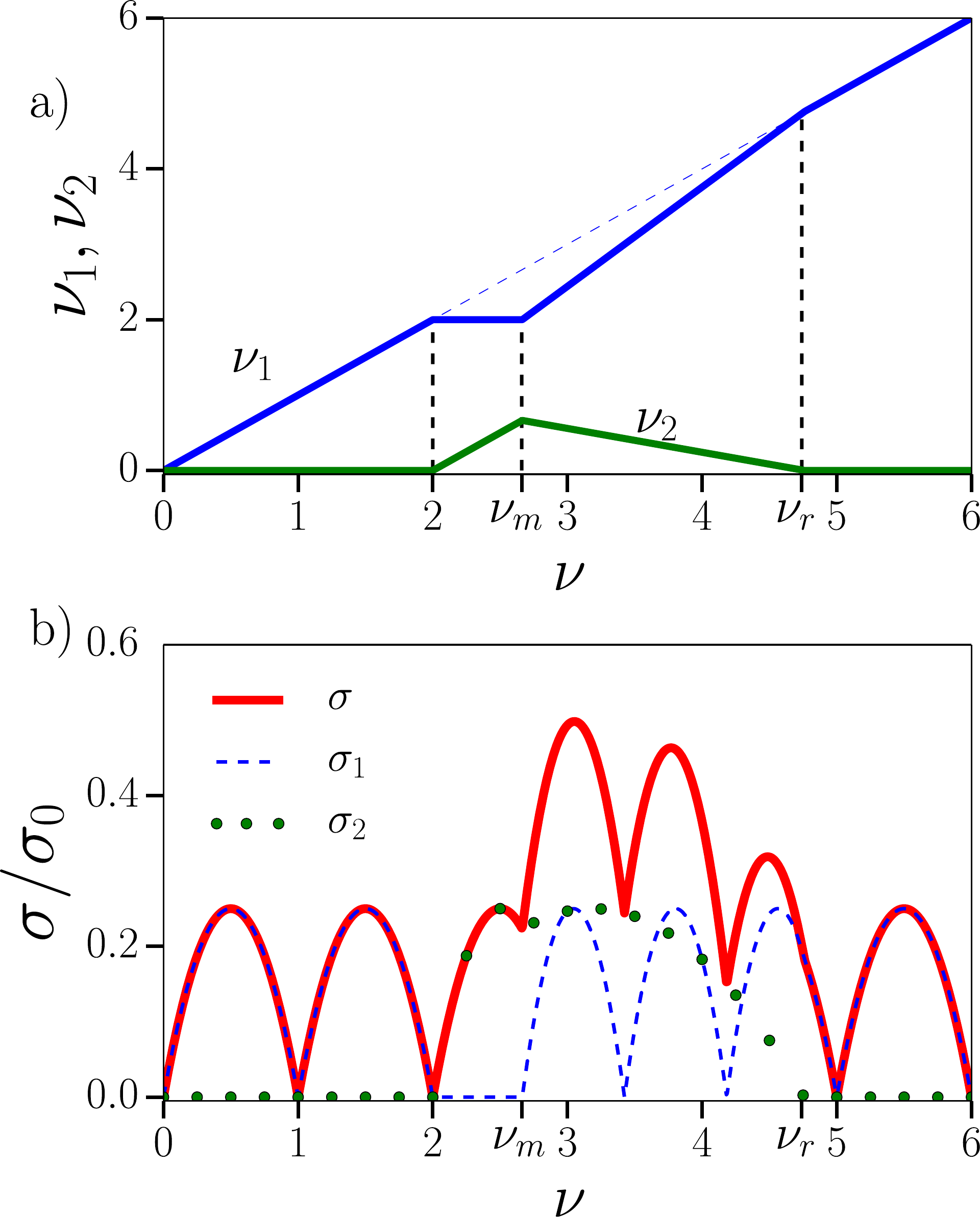}
\caption{(Color  online) The electron distribution and the surface conductivity of FET as obtained by  toy model of Sec. \ref{sec:analytic}. The ratio, $\Delta$ of the $1S-1P$ gap energy to the charging energy $e^2/\kappa D$ is equal to $6$, $D'=1.2D$. a) The  dependence of average numbers of electrons per NC  in the first   and second  layers $\nu_1$ and $\nu_2$  on the dimensionless surface density of electrons $\nu=\nu_1+\nu_2$.  The dashed line $\nu_1=\nu$ is plotted for clarity. $\nu_m$ and $\nu_r$ are critical points, where $\nu_1(\nu),\nu_2(\nu)$ change their  behaviour abruptly. b) The surface hopping conductivity  of FET based on NC array, $\sigma$, as a function of $\nu$. Here $\sigma_1,\sigma_2$ are conductivities  of the first and second layers. }
\label{fig:full}
\end{figure}

 Let us define average numbers of electrons per NC  in the first and the second layers (filling factors) as $\nu_1$ and  $\nu_2$  respectively, and introduce the dimensionless total surface density (total filling factor) $\nu=\nu_1+\nu_2$ .  With growing gate voltage or, in other words, growing $\nu$, at first, electrons occupy $1S$ levels of first layer NCs, so that $\nu_1=\nu$. When $\nu_1=\nu=2$ ($1S$ levels are totally filled) the $1S-1P$  gap induces occupation of $1S$ levels of  second layer NCs (see Fig.~\ref{fig:full}a). As a result, the first layer conductivity $\sigma_1$ vanishes while the second layer conductivity $\sigma_2$ starts to grow with $\nu$ (see Fig.~\ref{fig:full}b).  At a larger $\nu=\nu_m$ new electrons together with electrons  returning back from $1S$ levels of  second layer NCs   start filling $1P$ levels of  first layer NCs. Conductivity of the first layer $\sigma_1$ increases, and conductivity of the second layer $\sigma_2$ decreases and eventually vanishes. At   larger $\nu = \nu_r$ all the electrons of the second layer return to the first one, so that $\nu_2 = 0$ and $\nu_1= \nu$ at $\nu > \nu_r$. As a result of this nontrivial interplay of  the  electron  concentrations of the first  two layers, the total surface  conductivity $\sigma$ non-monotonically depends on the gate voltage (see Fig.~\ref{fig:full}b). 

Physics of non-monotonious redistribution of the electron concentration  between the first and the second layers can be qualitatively explained, as a result of an evolution of the effective screening radius of the array with the density of states at the Fermi level. For $\nu < 2 $, the Fermi level is in  $1S$ levels of  first layer NCs . The density of states at the Fermi level is large and the screening radius is smaller than $D$. At $\nu=2$ the Fermi level moves to the $1S-1P$ gap of first layer NCs. The $1S-1P$ gap dominates all other characteristic energies and leads to small characteristic density of states and larger screening radius, so that electrons start populating  $1S$ levels of  second layer NCs instead of  $1P$ levels of first layer NCs. As a result, electrostatic energy for the first layer goes down with the gate voltage. One can say that with growing gate voltage the $1S-1P$ gap  creates a "capacitor" made by the negative charge of second layer of electrons and the positive charge of missing electrons of the first layer.  The potential of this capacitor grows with $\nu$ until at a critical point, $\nu=\nu_m $, it exceeds the $1S-1P$ gap, so that $1P$ levels of  first layer NCs arrive at the Fermi level. At $\nu> \nu_m $, $\nu_1$ resumes to grow and the density of states grows with it. As a result, the effective screening radius decreases and some of second layer electrons return back to the first layer. At larger critical point, $\nu = \nu_r$, all the electrons of the second layer return to the first one, so that $\nu_2 = 0$ and $\nu_1= \nu$ at $\nu > \nu_r$.

Peculiar behavior of FET conductivity happens when the number of carriers per NC $\nu$ is close to three. It is difficult to achieve such carrier concentration  using  standard insulator. But high-$\kappa$ dielectric like $\mathrm{HfO_2}$ can help to achieve this goal. Recent progress in electrolyte-gated transistors \cite{e_Lee_Zhang_Lodge_Frisbie_2013,Fujimoto_Awaga_2013} has made this goal possible by using large polyions, which do not penetrate inside NC array, but create large gating charge close to NC surface. One can imagine also  conventional ionic liquid gating of an array with the gaps between NCs sealed.

The remainder of this paper is organized as follows.  In Sec.~\ref{sec:model} we define the FET model to be studied and we describe  parameters of a NC and FET based on an array of NCs.  Sec.~\ref{sec:computer} describes our  numerical calculation of electron concentrations of the first and the second layers and surface hopping conductivity of FET.  Results are presented in Sec.~\ref{sec:results}, along with an  explanation of relation between electron concentration in the first and the second  layers and the conductivity of the FET. This explanation is  based on the analytical study of  a toy model, where disorder is very small. In the Sec.~\ref{sec:analytic} we use the same toy model  to calculate $\nu_1(\nu)$ and $\nu_2(\nu)$ and critical concentrations $\nu_m$ and $\nu_r$ analytically.  Finally,  in Sec.~\ref{sec:capacitance}  we discuss the manifestation of the peculiar distribution of electrons between layers in the capacitance of the FET. 

\section{\label{sec:model} Model of a FET based on a NC array} 
In order to calculate the distribution of electrons between layers of  a NC array  and  transport properties of FET, we first introduce parameters that characterize  an isolated NC.  We assume that  a NC is a sphere of diameter $D$. We  suppose that the electron wave function is close to zero at the NC surface, due to large confining potential  barriers created by the insulator matrix surrounding the NC. Under these conditions the kinetic energy $E_Q(n)$ of the $n$th  electron added to the NC in the parabolic band approximation is :
\begin{equation}
E_Q(n) = \frac{\hbar^2}{m D^2} \times \left\{ 
\begin{array}{llr}
0,&  n = 0 & \\
19.74,&  n = 1, 2 &(1S)\\
40.38,&3 \leq n \leq 8&(1P)\\\dots\\
\end{array}
\right.
\label{eq:EQ}.  
\end{equation} 
As a result, the first two electrons added to the NC fill its $1S$ level, the next six ones fill its $1P$ level, and so on. 

The  kinetic energy of electrons is only a part of the total energy of the NC. One should add to it the total Coulomb energy of  electrons and their interactions. In general,  calculating the total Coulomb energy (self-energy) of the system is a difficult problem.  For our case, however, a significant simplification is available because the internal dielectric constant $\kappa_{NC}$ typically is much larger than  the external dielectric constant $\kappa$ of the insulator in which the NC is embedded. Specifically, the large internal dielectric constant $\kappa_{NC}$ implies that any internal charge $q$ is essentially completely compensated by the NC dielectric response, which leads to  homogeneous redistribution of the majority of the charge, $q (\kappa_{NC} - \kappa)/\kappa_{NC}$, over the NC surface.  In this way a semiconductor NC  can be considered as a metallic one in terms of its Coulomb interactions; namely, total Coulomb energy (self energy) is equal to $q^2/\kappa D$.

The  gap $\delta E=E_Q(3)-E_Q(2)$ between $1S$ and $1P$ levels of a NC and the energy necessary for adding one electron to  a neutral NC:
\begin{equation}
  \label{eq:charging}
  E_c= \frac{e^2}{\kappa D}
\end{equation}
, which we call charging energy, are the two most important energies of our theory. Below we use their ratio
\begin{equation}
\Delta = \frac{\delta E}{E_c} = \frac{20.64 \kappa \hbar^2}{me^2D } = 20.64 \frac{\kappa}{\kappa_{NC}} \frac{a_B}{D},  
\label{parameter}
\end{equation}
which grows with decreasing NC diameter $D$. 
Let us estimate $\Delta$. Consider, for example,  CdSe NC  with $\kappa_{NC} \simeq 9.2$ \cite{978-3-642-62332-5} and $D = 5$ nm surrounded by media with $\kappa \simeq 3.2 \ll \kappa_{NC}$ (see below). Using $m \simeq 0.12~m_e$ ($m_e$ is the electron mass) we have $E_c \simeq 0.08$ eV and $\Delta \simeq 5.7$. 

Now let us discuss characteristics of a weakly $n$-type doped array of such NCs. We assume that the number of donors per NC is much smaller than unity, but larger than the concentration of surface states or acceptors so that the Fermi level in the bulk of array is at  $1S$ levels  of NCs. We consider the case that NCs form a three dimensional cubic lattice structure with lattice constant $D'$ (see Fig.~\ref{fig:FET}). Usually each NC is coated by a thin layer of insulating ligands that maintain a separation  $D' - D$ between NCs.  We consider the case of finite array of $N=L \times L \times L_z$ NCs,  where integer $L,L,L_z$ are dimensions of the system along $x,y,z$ axises in  units of $D'$, respectively. $x$ and $y$ axises are parallel to the gate.

It should be noted that the effective dielectric constant $\kappa$ of the NC array is not simply equal to the dielectric constant  $\kappa_i$ of the insulator in which the NCs are embedded, but also includes the effect of polarization of NCs in response to an applied field.  This polarization effectively decreases both the Coulomb self-energy of a single NC and the interaction between neighboring NCs.  Generally speaking, the renormalization of the dielectric constant is not very strong, so that $\kappa$ is not very different from $\kappa_i$ even when $\kappa_{NC} \gg \kappa_i$.  The canonical Maxwell-Garnett formula gives the approximate relation \cite{Maxwell_treatise,PhysRevB.48.14936}:
\begin{equation}
  \label{eq:Maxwell}
\kappa \simeq \kappa_i \frac{\kappa_{NC} + 2 \kappa_i + 2 f(\kappa_{NC} - \kappa_i)}{\kappa_{NC} + 2 \kappa_i - f (\kappa_{NC} - \kappa_i)},
\end{equation}
where $f = \pi D^3/[6 (D')^3]$ is the volume fraction occupied by the NCs; for $f < 0.4$, this expression is accurate to within 8\% \cite{Doyle}.  As an example, for the case of a cubic lattice with $D = 5$ nm and $D' = 6$ nm (so that $f = 0.3$), $\kappa_{NC} \simeq 9.2$ and  $k_i=2$, the Maxwell-Garnett formula gives then $\kappa \simeq 3.2 \ll \kappa_{NC}$. 

In the array of NCs, an electron can  tunnel between NCs through insulating ligands. The hopping overlap integral  of electron states of neighboring NCs $t$ depends on the distance between nearest neighbor NCs as $\exp[(D-D'/a)]$, where $a \simeq 0.1-0.2~nm$ is the decay length of the electron wavefunction outside of the NC. The maximum value of $t$ can be achieved for the case of touching NCs. In the case of no disorder, electrons in the array of NCs can  be delocalized over an array. However, for a NC array there are two types of disorder: diagonal and non-diagonal, that prevent such behaviour.

Typically, standard deviation of diameter $D$ is 5\%\cite{Murray-AnnuRev30-2000}. It results in the random shift of  levels in a NC (diagonal disorder), which can be estimated as $W \simeq 2 \chi \delta E$, where $\chi \sim 0.05$ is relative standard deviation of diameter. For CdSe, $W \simeq 0.6 E_c \simeq 0.05$ eV.  In addition to energy disorder in every NC, there are fluctuations of distances between two nearest neighbor NCs (non-diagonal disorder). One can assume that a standard deviation of distance between two NCs is approximately equal to $\chi D$,  determined by the size variation of NCs. 

Typically, diagonal disorder alone leads to localization.  Indeed, it is well known that if $W$ is larger than the critical magnitude $W_c$ all states are localized~\cite{3540129952}. The critical parameter $W_c$ depends on $t$ and for a cubic lattice $W_c \simeq 8t$~\cite{3540129952}. The parameter $t$ was calculated for spherical NCs in the Ref. \onlinecite{Shabaev_Efros_Efros_2013}. For CdSe NCs with diameter $D \simeq 5$~nm, the maximum overlap integral for touching NCs ($D=D'$) $t \simeq 4$~meV. In that case $W_c \simeq 32$~meV, while $W \simeq 50$~meV. In other words,  even for the unlikely case of $D=D'$ all electron states are localized. 

For a NC array with nonzero $D'-D$ this means that every NC $i$ has an integer number of electrons $n_i$, and we can use Eqs. \eqref{eq:EQ},~\eqref{eq:charging},~\eqref{eq:Maxwell} to calculate the kinetic energy and the Coulomb self-energy of NCs. Thus, the Hamiltonian of the system can be written as
\begin{eqnarray}
H  & = &\sum \limits_{i=1}^N \frac{e^2 n_{i}^2 }{\kappa D} + \sum \limits_{i=1}^N \sum_{l = 0}^{n_{i}}E_Q(l) + \sum \limits_{i=1}^N n_{i}W_{i} + \nonumber \\
& & + \sum \limits_{\langle i,j \rangle }^{N} \frac{e^2  n_{i} n_{j}}{\kappa r_{i,j}} + 2\pi\nu \frac{e}{\kappa D'^2} \sum \limits_{k=2}^{N/L^2} (k-1) D' e \nu_k L^2.
\label{eq:H}
\end{eqnarray}
The first term of Hamiltonian is the Coulomb self-energy of  $i$-th NC. The second term describes the total quantum energy of  its $n_{i}$ electrons. The third term provides simplified description of disorder with the help of  a random shift energy of all levels of  $i$-th NC.  We assume that $W_i$ is distributed uniformly between $-W$ and $W$.  The  fourth term  is responsible for the Coulomb interaction between different NCs. The summation is over all NCs $i$ and $j$ except $i=j$. The last term is the potential energy of  electrons in the field of the gate  $2\pi\nu e/ \kappa D'^2$. Here,  $\nu$ is the dimensionless total surface density of electrons $\nu=\sum_i n_i /L^2=\sum_k \nu_k$, while $\nu_k$ is the number of electrons per NC in the $k$-th layer.  

The ground state for a particular system is defined by the set of electron occupation numbers $\{n_{i}\}$ that minimizes the Hamiltonian $H$.  Once one knows occupation numbers $\{n_{i}\}$ in the ground state one can determine $\nu_k$ and conductivity.

When the gate field induces electrons, most of them occupy only the first or the second layer of a  NC array that are close to the gate (see Fig.~\ref{fig:FET}). Indeed, one can use Debye-H\"uckel theory to estimate the screening radius $r_D=\sqrt{\kappa/4\pi e^2 g }$, where  $g$ is the density of states  on the Fermi level in a NC. If $\Delta=0$ then  the charging energy is the only characteristic energy, $g \sim 1/E_c D'^3$ and $r_D \simeq D'/2\sqrt{\pi}$. In that case, $r_D<D'$ and  electrons occupy only first layer NCs.  For large $\Delta$, the $1S$-$1P$ gap is  the other characteristic energy  and $r_D$ increases as $D'\sqrt{\Delta}/2\sqrt{\pi}$ but for realistic $\Delta \ll 20 $ all electrons occupy only  first or second layer  NCs. In that case, $\nu_1$ and $\nu_2$  are related to the gate  voltage $V$:

\begin{equation}
  \label{eq:nu_voltage}
\nu_1+\nu_2 \left(1+\frac{D'}{d}\right) = \frac{V \kappa}{4\pi d (e/D'^2)},
\end{equation}
where $d$ is the distance between first layer NCs and the gate (see Fig.~\ref{fig:FET}). Below we  study  $\nu_{1,2}(\nu),\sigma_{1,2}(\nu),\sigma(\nu)$ as a function of $\nu=\nu_1+\nu_2$.  In  experiments these quantities are studied as a  function of $V$. One can see that if $d \gg D'$ then $\nu=V \kappa/4\pi d (e/D'^2)$. For smaller $d$,  dependencies $\nu_{1,2}(V),\sigma_{1,2}(V),\sigma(V)$ on $V$ look like somewhat linearly stretched    $\nu_{1,2}(\nu),\sigma_{1,2}(\nu),\sigma(\nu)$.

According to the above Debye-H\"uckel approach, $\nu_1(\nu)$ and $\nu_2(\nu)$  monotonically increase with $\nu$. We show below that actually in a NC array $\nu_1$ and $\nu_2$ do not change  monotonically. This leads to a peculiar nonmonotonic change of surface conductivity of induced electrons (see Figs.~\ref{fig:full} and \ref{fig:numerical}).

\section{Computer simulation} \label{sec:computer}

In this section we describe our computational method for calculating $\nu_1,\nu_2,\sigma$ of a NC array  at a given value of $\nu$, $T$ and $D$. A reader  interested in the result only, can go to the next section. These calculations are based on a computer simulation of a finite  array of $L \times L \times L_z$ NCs, with  periodic boundary condition along parallel to the gate $x$ and $y$ axises. We use the  procedure outlined in Refs. \onlinecite{Conductivity_Brian_Shklovskii}, \onlinecite{Reich_Efros_Shklovskii}. First, we specify the total number of electrons in our system $M$ and $\nu=M/L^2$. The initial values of  the electron numbers $\{n_{i}^0\}$ are then assigned randomly in such a way that  $\sum_{i} n_{i}^0 = M$. The simulation then assigns a random energy shift $W_i$ at each NC $i$. The program then searches for the ground state by looping over all NC pairs $ij$ and attempting to move one electron from NC $ i $ to $j $. If the move lowers the Hamiltonian $H$, then it is accepted, otherwise it is rejected. In this way we arrive to a pseudogrond state, which describes a NC array at low temperatures \cite{Conductivity_Brian_Shklovskii}. As soon as this state is found one can calculate the conductivity of the system $\sigma$. 

In order to compute $\sigma$ one can determine the  highest filled electron level, $\varepsilon_{i}^{(f)}$, and the lowest empty electron level, $\varepsilon_{i}^{(e)}$, at each NC $i$ at a given electron distribution $\{n_{i}\}$.  Specifically,
\begin{eqnarray} 
\varepsilon_{i}^{(f)} & = & E_Q(n_{i}) +W_i + \frac{e^2[ n_{i}^2 - (n_{i} - 1)^2]}{\kappa D} \nonumber \\
& & + \sum_{j \neq i} \frac{en_{j}}{\kappa r_{ij}}+2\pi \nu \frac{e}{\kappa D'^2}  k D'
\label{eq:enf}
\end{eqnarray}
and
\begin{eqnarray} 
\varepsilon_{i}^{(e)} & = & E_Q(n_{i} + 1) +W_i+ \frac{e^2[(n_{i} + 1)^2 - n_{i}^2]}{\kappa D} \nonumber \\
& & +\sum_{j \neq i} \frac{e n_{j}}{\kappa r_{ij}} +2\pi \nu \frac{e}{\kappa D'^2}  k D'.
\label{eq:ene}
\end{eqnarray}

In the results presented below we define electron energies $\varepsilon$ relative to the Fermi level $\mu$, which is calculated for each realization of the simulation as $\mu = [ \min\{ \varepsilon_i^{(e)} \} - \max\{ \varepsilon_i^{(f)} \}]/2$.  In this way $\varepsilon < 0$ corresponds to filled electron states $\varepsilon^{(f)}$ while $\varepsilon > 0$ corresponds to empty states $\varepsilon^{(e)}$. Once the pseudo-ground state energies $\{\varepsilon_{i}^{(f)} \}$ and $\{\varepsilon_{j}^{(e)}\}$ are determined, we calculate the hopping conductivity of the system by mapping the simulated NC array to an effective Miller-Abrahams resistor network. To calculate conductance $\sigma_{ij}$ between NCs $i$, $j$ we consider only electron transfer among the highest filled states, $\varepsilon^{(f)}$, and the lowest empty states, $\varepsilon^{(e)}$. This is appropriate when the temperature is small enough that $k_B T \ll e^2/\kappa D$, so that thermal excitation of multi-electron transitions is exponentially unlikely.

Since each NC has two energy levels that can participate in conduction, $\varepsilon^{(f)}$ and $\varepsilon^{(e)}$, one can say that there are four parallel conduction processes that contribute to the conductance between two NCs $i$ and $j$: one for each combination of the initial energy level at site $i$ (either $\varepsilon_{i}^{(f)}$ or $\varepsilon_{i}^{(e)}$) and the final energy level at site $j$ (either $\varepsilon_{j}^{(f)}$ or $\varepsilon_{j}^{(e)}$).  Each of these four processes has a corresponding effective conductance $\sigma_{ij}^{(\alpha \beta)}$, where $\alpha, \beta = f, e$
\begin{equation}
  \label{eq:conductance}
  \sigma^{(\alpha \beta)}_{ij}=\sigma_0 \exp \left( - \frac{\varepsilon_{ij}^{(\alpha, \beta)}}{k_B T} \right) ,
\end{equation}
where  $\sigma_0$ is a prefactor that have a weak dependence on temperature and following Ref. \onlinecite{3540129952}:
\begin{equation}
\varepsilon_{ij}^{(\alpha, \beta)} = \left\{
\begin{array}{lr}
|\varepsilon_{j}^{(\beta)} - \varepsilon_{i}^{(\alpha)}| - \frac{e^2}{\kappa r_{ij}}, &  \varepsilon_{j}^{(\beta)}\varepsilon_{i}^{(\alpha)} < 0 \vspace{2mm} \\
\max \left[ \left|\varepsilon_{j}^{(\alpha)} \right|, \left|\varepsilon_{j}^{(\beta)} \right| \right], &  \varepsilon_{j}^{(\beta)}\varepsilon_{i}^{(\alpha)} > 0
\end{array}
\right.
.
\label{eq:epsij}
\end{equation}
These four conductances can be said to be connected in parallel between NCs $i$ and $j$, and the condactance between two NC $i$ and $j$ is the sum of these $\sigma_{ij}$ conductances.

If one knows conductance between any two NC, one can determine $\sigma$ by using Kirchhoff law. First, we assign  a  voltage $V_0 x/L$ to every NC with coordinate $x,y,z$  in the array. We use Gauss-Seidel method to find relaxed voltage distribution in steady state. Specifically, the program  loops through all NCs and determines the voltage on the NC so that current that flows into NC is equal to the current that flow from NC, namely, we find voltage $V_j$ on  every NC $j$ from condition :
$$
\sum_i \sigma_{ij} (V_{i}-V_{j})=0,
$$
Aiming at relatively high $T$, we keep only NCs $i$, which are nearest neighbors (NN)  of NC $j$.  The voltages $V(x=0)=0$, $V(x=L)=V_0$ are fixed. After each iteration, we  calculate  current through the system $I$.

We stop the iterating process when current  starts changing by less than  0.1\%. After that one can calculate the surface (2D) conductivity of the system which, of course, doesn't depend on $L_z$.

\begin{figure}
\includegraphics[width=1\linewidth]{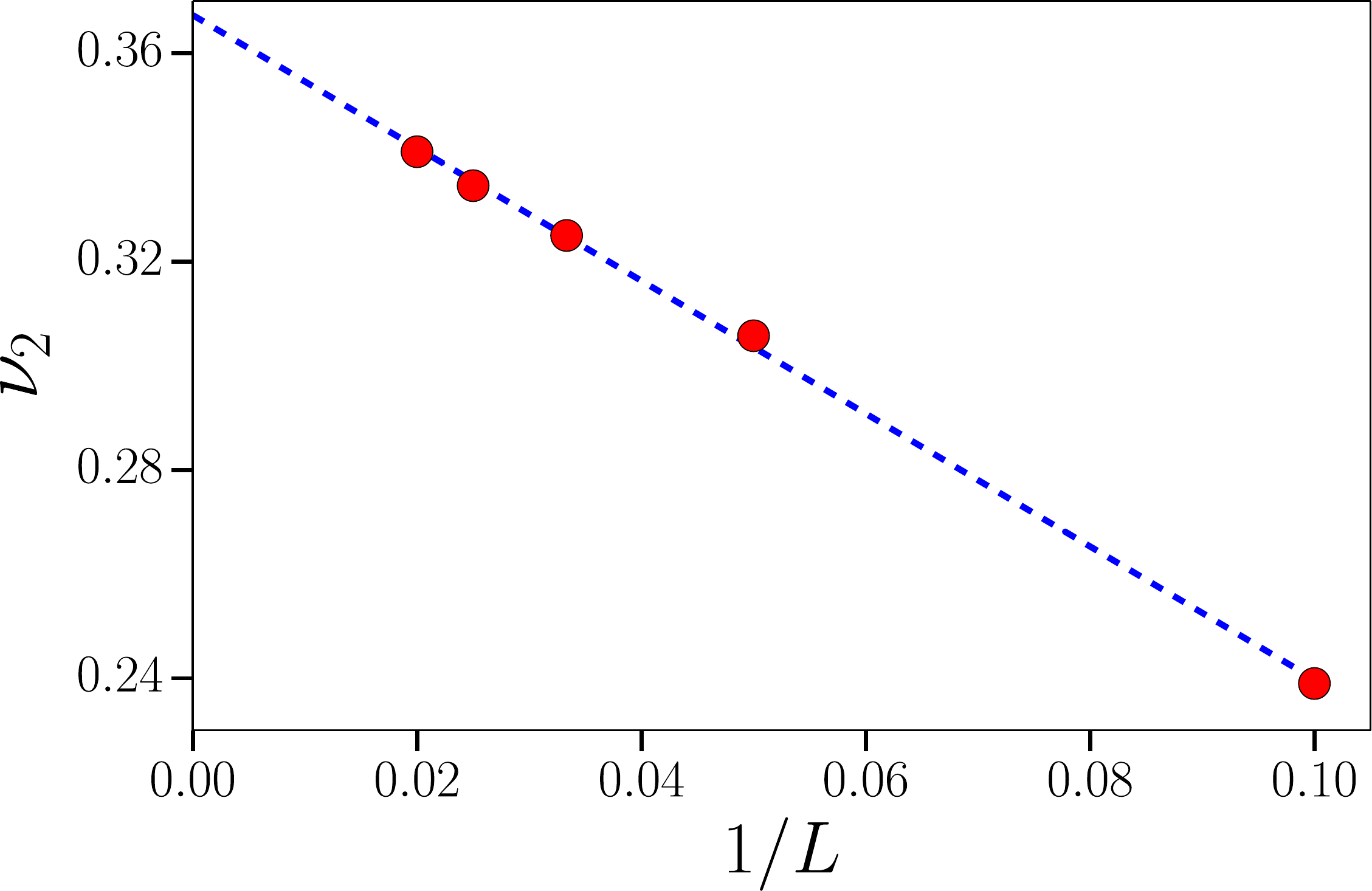}
\caption{(Color  online) The dependence of $\nu_2$  on the size of the NC array  $L$ at $\nu=2.5$. Numerical results are shown by circles. The size of points reflects a  computation error bars. The linear   dependence of  $\nu_2$ on $1/L$ is used to extrapolate to $L=\infty$.}
\label{fig:example_MC}
\end{figure}

The number of electrons on the first layer $\nu_1$ and the second layer $\nu_2$   in the pseudoground state and the conductivity of FET $\sigma$ are averaged over 10  different realizations of  random  sets of the electron numbers $\{n_{ik}^0\}$. In order to study the size dependence of these parameters, we vary  $L$ from $10$ to $50$ and keep $L_z=5$. All results below are obtained by extrapolation to $L=\infty$ . For example, one can see on Fig. \ref{fig:example_MC} an extrapolation of $\nu_2$ at $\nu=2.5$.

\section{Results and discussion} \label{sec:results}

\begin{figure}
\includegraphics[width=1\linewidth]{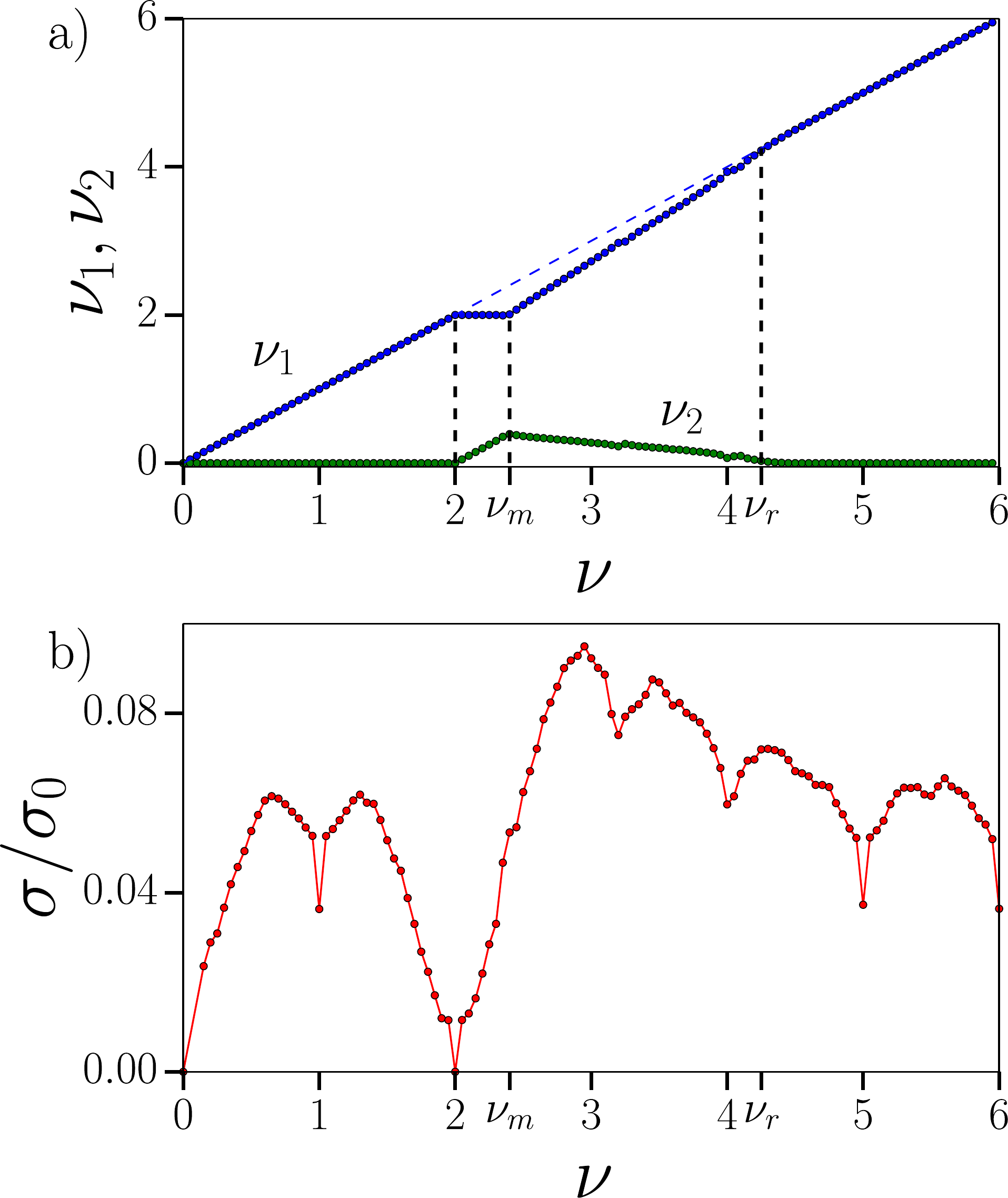}
\caption{(Color  online) The numerical results for FET model. $1S-1P$ gap is equal to $6E_c$, $T=0.2 E_c$, $W=0.5 E_c$, $D'=1.2D$. a)  The  average number of electrons per NC in the first, $\nu_1(\nu)$, and second layer, $\nu_2(\nu)$. The size of points reflects a computation error bars. The dotted line $\nu_1=\nu$ is plotted for clarity. b) The hopping conductivity  of FET, $\sigma$, as a function of $\nu$. The solid line is a guide to the eye. }
\label{fig:numerical}
\end{figure}

All the numerical results for the electron distribution and the conductivity are obtained for $\Delta=6$, $W=0.5 E_c$ and $D'=1.2 D$. In Fig.~\ref{fig:numerical}a we show results of the numerical simulation for $\nu_1(\nu),\nu_2(\nu)$ at $\nu < 6$  in the ground state ($T=0$). Dependencies of $\nu_1(\nu)$ and $\nu_2(\nu)$ are  different in four ranges of $\nu$.  In the first range,  $\nu \le 2$, all electrons settle in the first layer, $\nu_1=\nu$ and $\nu_2=0$.  For $2\le\nu\le\nu_m$,  first layer NCs have exactly 2 electrons $\nu_1=2$, while  $\nu_2$ increases linearly with $\nu$ until it reaches a maximum at $\nu=\nu_m$. In the range $\nu_m\le\nu\le\nu_r$, while  $\nu_1$  increases and $\nu_2$ decreases with  $\nu$ and vanishes at $\nu=\nu_r$. In the last range $\nu>\nu_r$, again $\nu_1=\nu$  and $\nu_2=0$. 

In Fig.~\ref{fig:numerical}b we show  numerical results for $\sigma(\nu)$ for  $ T=0.2 E_c/k_B$. As  mentioned above $E_c$ for CdSe NC with $D=5$ nm is close to $80$ meV, so that $ T=0.2 E_c/k_B \simeq 190$ K. We see that conductivity $\sigma(\nu)$ has a deep minimum at $\nu = 2$, smaller minima at $\nu =1, 4, 5, 6$ and a weaker minimum at $\nu \simeq 3.2$. There is also a large peak of conductivity at $\nu \simeq 2.9$. It is $55 \%$ higher than peaks at $\nu=0.5$ and $1.5$

In order to explain the origin of this peculiar behaviour of $\sigma(\nu)$, below we study an analytically tractable   toy model, where disorder is very small, $W \ll k_B T \ll E_c$, (but large enough to provide electron localization). We find analytical dependencies $\nu_1(\nu)$, $\nu_2(\nu)$ in the Sec. \ref{sec:analytic} . They are shown in Fig.~\ref{fig:full}a for $\Delta=6$. One can see that two figures Fig.~\ref{fig:full}a and Fig.~\ref{fig:numerical}a  are similar to each other. For smaller $\Delta$ both critical points $\nu_m$, $\nu_r$ tend to $2$ so that the plateau of $\nu_1(\nu)$ curve  shrinks and  vanishes at $\Delta=2.5$. This is in qualitative agreement to what happens with Fig.~\ref{fig:numerical}a for numerical modeling of more realistic disordered model.

Below we  concentrate on the behavior of $\sigma(\nu)$ . In our toy model,  all conductances between two NN NCs $i$ and $j$, $\sigma_{ij}$, within a given layer are identical. Therefore, there are identical constant gradient of  electro-chemical potential  along both the first and the second layers and there are no interlayer currents between NN NCs of the two layers. As a result, the conductivity of FET $\sigma$ is the sum of conductivities of two independent layers of a NC array  $\sigma_1$ and $\sigma_2$. At the same time, each of conductivities $\sigma_1, \sigma_2$ is equal to the conductance between NN NCs of the corresponding layer. 

Let us consider, for example, the first layer in the range of filling factors $ 0 < \nu_1 < 1$, when  each NC is occupied by an electron with probability  $ \nu_1$ and  is empty with  probability $(1-\nu_1)$.  At $T \ll E_c$,  probability of an electron hopping to  an occupied NN NC is proportional to $\exp(-E_c/T) \ll 1$ and can be ignored. Thus, only hops between an occupied and empty NN NCs contribute to the conductivity. The conductivity is proportional to the number of occupied and the number of empty NCs, i.e. $\sigma_1 (\nu_1) = \sigma_0 \nu_1 (1-\nu_1)$. For $1< \nu_1< 2$ all NCs contain at least one electron, while some of them contain two electrons. Second electrons  hop between  NCs with one electron. Repeating the same argument we get $\sigma_1(\nu_1) =\sigma_0 (\nu_1-1) (2-\nu_1)$ for $1< \nu_1< 2$. We can use similar expressions for the conductivity of the first layer at $\nu_1> 2$. 
We  can also write similar expressions for  $\sigma_2 (\nu_2)$. Using  functions $\nu_1 (\nu)$ and $\nu_2 (\nu)$ calculated in the next section and  shown for $\Delta=6$ on  Fig.~\ref{fig:full}a, we arrive to a layer conductivity dependencies $\sigma_1 (\nu_1(\nu))$, $\sigma_2 (\nu_2(\nu))$, which are shown on  Fig.~\ref{fig:full}b. Combining both $\sigma_1,\sigma_2$ we get  the total conductivity of FET: 
\begin{equation}
  \label{eq:sigma_total}
\sigma (\nu) = \sigma_1 (\nu_1 (\nu))  + \sigma_2 (\nu_2 (\nu)),
\end{equation}
which is plotted in  Fig.~\ref{fig:full}b as well. This result reflects non-trivial redistribution of electrons between layers. Minima at $\nu=1,2,5,6$ correspond to occupation of the first layer NCs with equal integer number of electrons. Indeed, when $0<\nu<2$, the conductivity is determined by the first layer. In the range of $2<\nu<\nu_m$ all  first layer NCs contain two electrons, while the second layer hosts the rest $\nu-2$ electrons per NC. As a result the first layer does not contribute to the conductivity and only the second layer conducts. In the next range $\nu_m<\nu<\nu_r$ both the first layer and the second layers conduct. For larger  $\nu>\nu_r$ the second layer is empty and does not conduct. As a  result, the total conductivity peaks  in the range of gate voltage where $\nu_m<\nu<\nu_r$. 

Now we can return to the discussion of shown in Fig. \ref{fig:numerical}b conductivity on $\nu$ for more realistic model with disorder. Just as in the toy model in the range $\nu<2$ and $\nu>\nu_r$ only the first layer contribute to the conductivity.  Minima at $\nu=1,2,4,5,6$ correspond to occupation of the first layer NCs with equal integer number of electrons. Because the activation energy  at $\nu=2$  is larger than activation energy at other points, the minimum at $\nu=2$ is deeper than at other points. In the range $\nu_m<\nu<\nu_r$ both the first and the second   layers of a NC array contain electrons and both conduct. This leads to the peak in the conductivity of the FET.

\section{Toy model theory of screening}
\label{sec:analytic}

In order to understand our numerical results let us evaluate analytically $\nu_1$ and  $\nu_2$  using Hamiltonian $H$ for the case without disorder, i.e., $W=0$. We start from the condition at which all the electrons occupy the first layer. In order to do this, we consider the case when all first layer NCs (except one NC $A$)  contain integer number of electrons $\nu_1=n_{i1}=\nu$ ($i \ne  A$), one NC $A$ has one additional electron $n_{A1}=\nu+1$ electrons, all second layer NCs do not contain electrons at all, namely, $n_{i2}=\nu_2=0$. The energy of such an electron configuration ``1'', $H_1$, can be calculated using Hamiltonian \eqref{eq:H}. We can compare the energy of this electron configuration with the energy of configuration ``2'', $H_2$, in which the electron moves from the first layer NC $A$ to  any  second layer NC. The energy difference $ E_{21}=H_2 - H_1$ between two electron configurations ``1'' and ``2'' can be readily derived:
 
 \begin{equation}
E_{21}(\nu) = \frac{e^2}{\kappa D}\left( \frac{D}{D'}\left[ 2\pi \nu - \nu \alpha   \right ]  -2\nu - \Delta \Theta(\nu+\delta-2) \right ).
 \label{eq:difference}
 \end{equation}

The first two terms $E_{21}(\nu)$ in the bracket $\left[ \cdots \right]$ describe effects of the Coulomb interaction of moved electron. The first one is the energy change of the moved electron in the electric field of the gate.  The second one describes  difference in the Coulomb interaction of the moved electron with the rest of the electron system. In electron configuration ``1''  this energy is $$ \nu \frac{e^2}{\kappa D'}  \sideset{}{'} \sum   \limits_{i,j=-\infty}^{\infty}  \left( \frac{1}{\sqrt{i^2+j^2}} \right),$$ where the summation is  over all $i,j$ except $i=j=0$. In  the electron configuration ``2'' this energy is  $$\nu \frac{e^2}{\kappa D'}  \sum \limits_{i,j=-\infty}^{\infty} \left( \frac{1}{\sqrt{i^2+j^2+1}} \right).$$ 
$$
  \alpha= \sideset{}{'}\sum \limits_{i,j=-\infty}^{\infty} \left( \frac{1}{\sqrt{i^2+j^2}} -  \frac{1}{\sqrt{i^2+j^2+1}} \right) -1 \simeq 2.37
$$
The third term describes a change of the self energy. In the  electron configuration ``1'' the self energy is equal to $(\nu+1)^2 e^2/\kappa D $ and in the electron configuration ``2'' the self-energy is equal to  $(\nu^2+1) e^2/\kappa D $. The last term is related to the loss of  quantum energy, because the electron  moves from the $1P$ state of the NC A to the $1S$ state of the second layer NC. $\Theta(x)$ is Heaviside function, one electron on the second layer gives negligible small correction to the $\nu$:  $\delta>0$. We do not take into account this number in other terms, but in the   $\Theta(\nu+\delta-2)$, it determines the value of Heaviside function for $\nu=2$. In the result, from \eqref{eq:difference} one can see that $\Delta$ and the charging energy drive an electron to second layer NCs and the electric field is screened by other electrons prevent this.

\begin{figure}
\includegraphics[width=1\linewidth]{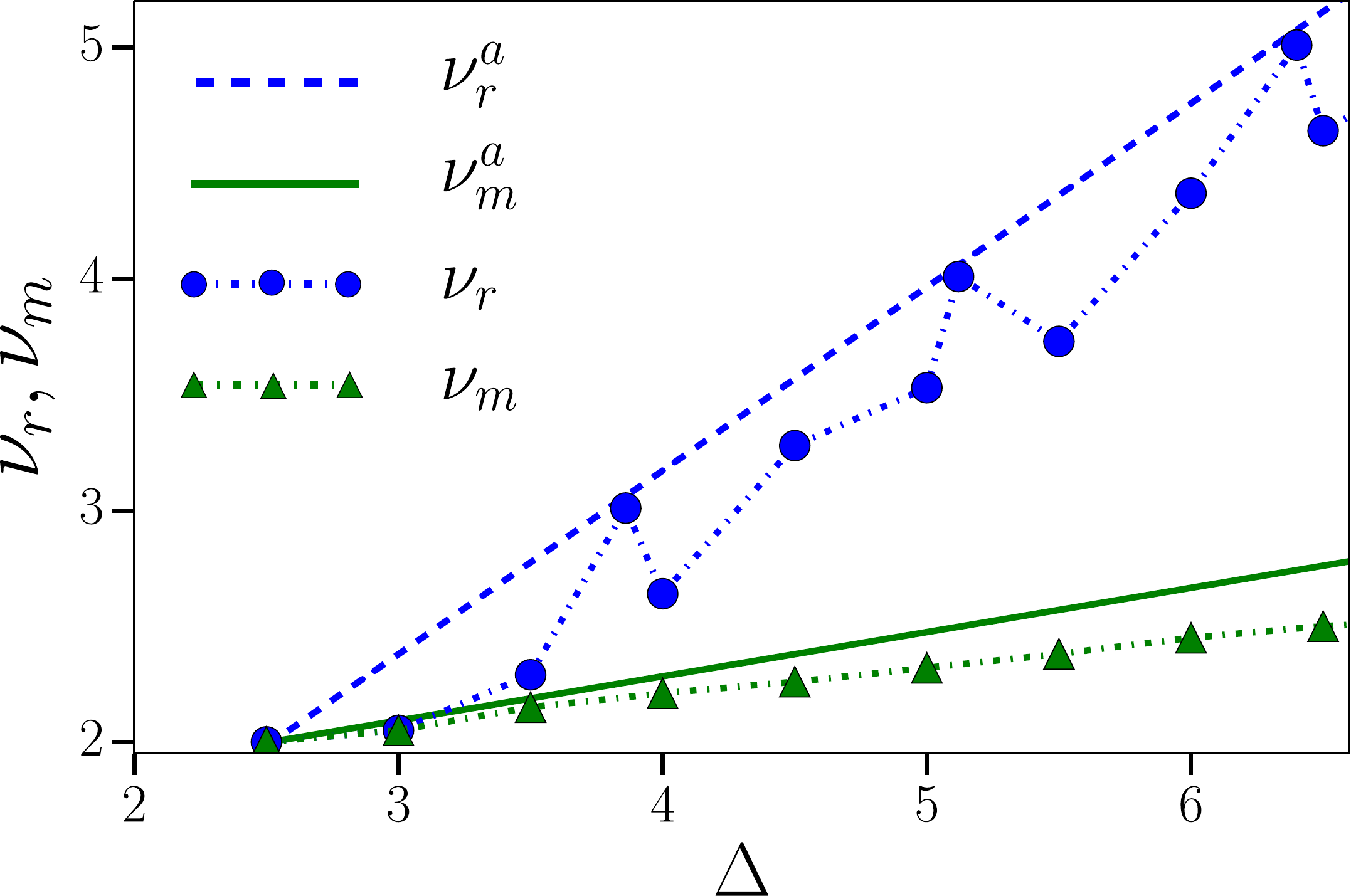}
\caption{(Color  online) Dependencies of critical points $\nu_m$ and $\nu_r$ (defined in Fig.~\ref{fig:full}) on $\Delta$ for $D'=1.2D$. Numerical results are shown by filled triangles and circles. The size of points reflects a  computation error bars. Dashed-dotted lines are a guide to the eye. The  analytical  dependencies   $\nu_{m}^{a}(\Delta)$ and $\nu_{r}^{a}(\Delta)$  given by Eqs. \eqref{eq:nu_m} and \eqref{eq:nu_v}  are shown by the solid and dashed lines correspondingly.}
\label{fig:numerical2}
\end{figure}

The $E_{21}(\nu)$ is derived for an integer number of electrons per NC in the first layer $\nu=\nu_1$, in that case this formula is an exact result. For noninteger $\nu=\nu_1$, one can think about $E_{21}(\nu)$ as the estimation for the energy difference.

In the range of $\nu$, where  $E_{21}(\nu) > 0$ all electrons occupy first layer NCs. This happens at $\nu < 2$ or $\nu> \nu_r$, where:
 
 \begin{equation}
\nu_r = \frac{\Delta}{\frac{D}{D'}\left[ 2\pi -  \alpha   \right ]  -2}.
\label{eq:nu_v}  
 \end{equation}
 
The dependence  $\nu_r$ on $\Delta$ is shown in Fig.~\ref{fig:numerical2}.  Small disagreement  between the numerical result $\nu_r$ and the analytical result $\nu^{a}_{r}$ can be explained by the creation of electron lattices in the first and the second layers at noninteger $\nu$, that has not been taken into account.  

If $\nu_r = 2$ all electrons always occupy first layer NCs. This happens if $\Delta<\Delta_c$:

\begin{equation}
\Delta_c = \frac{D}{D'}\left[ 4\pi -  2\alpha   \right ]  -4 
  \label{eq:critical_Delta}
\end{equation}

One can see that, if  $D' > 1.95 D$ the electrons redistribute between two layers even at  $\Delta=0$. For $D'=1.2D$, \eqref{eq:critical_Delta} gives $\Delta_c \simeq 2.5$ in close agreement with  numerical result $2.52 \pm 0.02$ (see Fig.~\ref{fig:numerical2}). If $\Delta>\Delta_c$ then there is a range of $2<\nu<\nu_r$, in which electrons coexist in two layers. In the range $2<\nu<\nu_m$,  every NC of the first layer contains two electrons, i.e., $\nu_1=2$ and in this range  $\nu_2$  linearly depends  on $\nu$:   $\nu_2=\nu-2$.  One can find  $\nu_m$  if one rewrites the energy differences $E_{12}$  Eq. \eqref{eq:difference} for the case when the number of electrons on the second layer is not equal to zero  $0<\nu_2 < 1$ and $\nu_1=2$:

\begin{equation}
E'_{12}=\frac{e^2}{\kappa D}\left( \frac{D}{D'}\left[ 2\pi \nu - 2 \alpha   \right ]  -4 - \Delta \right ).
 \label{eq:difference_2}
\end{equation}
One can see that with increasing  $\nu$ the interaction between additional  electron induced by the gate and the electric field increases (term $2\pi\nu$). When the gain energy $E'_{12}>0$ induced electrons again settle in the first layer and  the linear growth of $\nu_2$ stops. This condition defines boundary  $\nu_m$

\begin{equation}
  \label{eq:nu_m}
  \nu_m=\frac{\Delta + 4 + 2\alpha D/D' }{2\pi D/D'}.
\end{equation}

 The function  $\nu_m$  is shown in Fig.~\ref{fig:numerical2}. Again, small disagreement between the numerical result  $\nu_m$ and the analytical result $\nu_m^{a}$  can be explained by the creation of electron lattices at noninteger $\nu$, that  have not been take into account. 

In the range of $\nu_m<\nu<\nu_r$ electrons return back from second layer NCs  to  first layer NCs. Energy $E_{21}$ \eqref{eq:difference} depends linearly on $\nu$ and because of that  the dependence of $\nu_2(\nu)$ is linear one.  In the result, we show analytical dependencies $\nu_1(\nu)$ and $\nu_2(\nu)$ on Fig.~\ref{fig:full}a, and plot conductivity on Fig. \ref{fig:full}b  with recipe that is described in the previous  section.

\section{Capacitance of a NC based FET} \label{sec:capacitance}

The peculiar electron distribution between two layers of a NC array  can be tested by measurements of FET differential capacitance $$C=\frac{dQ}{dV}.$$ Here $Q=\nu S e/D'^2$ is the net charge induced in  a NC array with total surface area $S$ by voltage $V$. Using Eq. \eqref{eq:nu_voltage} we get $$C=\frac{S\kappa}{4\pi d^{*}},$$ where $$d^{*}=d+D' \frac{d\nu_2}{d\nu}.$$ In the range $\nu<2$, $\nu_2$ does not change : $d\nu_2/d\nu=0$, because of that  $d^{*}=d$ and the capacitance is equal to geometrical capacitance $C=C_g=S\kappa/4\pi d$. In the range $2<\nu<\nu_m$, $\nu_2$ increases with $\nu$ : $d\nu_2/d\nu=1$, because of that  $d^{*}=d+D'$ and $C<C_g$. In the range $\nu_m<\nu<\nu_r$, $\nu_2$ decreases with $\nu$ :  $d\nu_2/d\nu<0$, because of that  $d^{*}<d$ and  $C>C_g$. One can see from the Fig. \ref{fig:numerical}a that in the range $\nu_m<\nu<\nu_r$, for $\Delta=6$ and  $W=0.5 E_c$ $d\nu_2/d\nu \simeq -0.3$ i.e. $d^{*}=d-0.3D'$. In the range $\nu>\nu_r$, when all electrons are back from the second layer $\nu_2$ does not change :  $d\nu_2/d\nu=0$ and $d^{*}=d$, $C=C_g$.

\begin{acknowledgments}
The authors would like to thank B. Skinner, D. Talapin, D. Frisbie, D. Norris, M. Kang, M. Law  for helpful discussions. This work was supported primarily by the MRSEC Program of the National Science Foundation under Award Number DMR-0819885 and T. Chen  was  supported by the Louise T. Dosdall fellowship.  
\end{acknowledgments}

\end{document}